%
%
\input harvmac.tex
\def\cn{{\cal N}}
\def\IR{\relax{\rm I\kern-.18em R}}
\def\IZ{\relax\ifmmode\hbox{Z\kern-.4em Z}\else{Z\kern-.4em Z}\fi}
\def\co{{\cal O}}
\def\vev#1{\left\langle #1 \right\rangle}
\def\Im{{\rm Im}}
\def\tp{{\tilde\Phi}}
\def\tm{{\tilde m}}
\def\tu{{\tilde u}}
\def\tt{{\tilde \tau}}
\def\myTitle#1#2{\nopagenumbers\abstractfont\hsize=\hstitle\rightline{#1}%
\vskip 0.5in\centerline{\titlefont #2}\abstractfont\vskip .5in\pageno=0}
%
%
\def\np#1#2#3{{\it Nucl. Phys.} {\bf B#1} (#2) #3}
\def\pl#1#2#3{{\it Phys. Lett.} {\bf #1B} (#2) #3}
\def\plb#1#2#3{{\it Phys. Lett.} {\bf #1B} (#2) #3}

\def\prep#1#2#3{{\it Phys. Rept.} {\bf #1} (#2) #3}

\def\atmp#1#2#3{{\it Adv. Theor. Math. Phys.} {\bf #1} (#2) #3}
\def\jhep#1#2#3{{\it J. High Energy Phys.} {\bf #1} (#2) #3}
\def\ncim#1#2#3{{\it Nuovo Cim.} {\bf #1A} (#2) #3}
%
%

\lref\bagr{T. Banks and M. B. Green, ``Non-perturbative effects in 
$AdS_{5}\times  S^5$ string theory and $d = 4$ SUSY  Yang-Mills,'' 
hep-th/9804170, \jhep{\bf 9805}{1998}{002}.}

\lref\grgut{M. B. Green and M. Gutperle, 
``Effects of D-instantons,'' hep-th/9701093,
\np{498}{1997}{195}.}

\lref\holo{ M.~Bershadsky, S.~Cecotti, H.~Ooguri and C.~Vafa, 
``Holomorphic anomalies in topological field theories,'' hep-th/9302103,
\np{405}{1993}{279}.}

\lref\witm{ E. Witten, 
``Solutions of four-dimensional field theories via M-theory,''
hep-th/9703166, 
\np{500}{1997}{3}.}

\lref\mqcd{E. Witten, ``Branes and the dynamics of QCD,'' hep-th/9706109,
\np{507}{1997}{658}.}

\lref\mqcdoog{K.~Hori, H.~Ooguri and Y.~Oz,
``Strong coupling dynamics of four-dimensional N = 1 gauge 
theories from  M theory fivebrane,'' hep-th/9706082, \atmp{1}{1998}{1}.}

\lref\juan{J. M. Maldacena, ``The large $N$ limit of superconformal field
theories and supergravity,'' hep-th/9711200, \atmp{2}{1998}{231}.}

\lref\gkp{S. S. Gubser, I. R. Klebanov and A. M. Polyakov, 
``Gauge theory correlators
from non-critical string theory,'' hep-th/9802109,
\plb{428}{1998}{105}.}

\lref\wittenAdS{E. Witten, ``Anti-de-Sitter space and holography,''
hep-th/9802150, \atmp{2}{1998}{253}.}

\lref\magoo{O. Aharony, S. S. Gubser, J. M. Maldacena, H. Ooguri and
Y. Oz, ``Large $N$ field theories, string theory and gravity,''
hep-th/9905111, \prep{323}{2000}{183}.}

\lref\gubser{
S.~S.~Gubser,
``Curvature singularities: the good, the bad, and the naked,''
hep-th/0002160.}

\lref\ken{K.~Intriligator,
``Bonus symmetries of $\cn = 4$ super-Yang-Mills correlation functions 
via  $AdS$ duality,'', hep-th/9811047, \np{551}{1999}{575}.}

\lref\konishione{K. Konishi, ``Anomalous supersymmetry transformation
of some composite operators in SQCD,'' \pl{153}{1984}{439}.}

\lref\konishitwo{K. Konishi and J. Shizuya, ``Functional integral
approach to chiral anomalies in supersymmetric gauge theories,''
\ncim{90}{1985}{111}.}

\lref\gppz{L.~Girardello, M.~Petrini, M.~Porrati and A.~Zaffaroni,
``The supergravity dual of $\cn = 1$ super Yang-Mills theory,''
hep-th/9909047, \np{569}{2000}{451}.}

\lref\polstr{J. Polchinski and M. J. Strassler, ``The string dual of a
confining four-dimensional gauge theory,'' hep-th/0003136.}

\lref\vafwit{C.~Vafa and E.~Witten, ``A strong coupling test of S duality,''
hep-th/9408074, \np{431}{1994}{3}.}

\lref\koblitz{N. Koblitz, ``Introduction to Elliptic Curves and 
Modular Forms'' (Springer-Verlag, 1984).}

\lref\donwit{R.~Donagi and E.~Witten, ``Supersymmetric Yang-Mills
theory and integrable systems,'' hep-th/9510101,
\np{460}{1996}{299}.}

\lref\donwit{R.~Donagi and E.~Witten, ``Supersymmetric Yang-Mills
theory and integrable systems,'' hep-th/9510101,
\np{460}{1996}{299}.}

\lref\kssone{S.~Kachru, M.~Schulz and E.~Silverstein,
``Self-tuning flat domain walls in 5d gravity and string theory,''
hep-th/0001206.}

\lref\ksstwo{S.~Kachru, M.~Schulz and E.~Silverstein,
``Bounds on curved domain walls in 5d gravity,''
hep-th/0002121.}

\lref\adks{N.~Arkani-Hamed, S.~Dimopoulos, N.~Kaloper and R.~Sundrum,
``A small cosmological constant from a large extra dimension,''
hep-th/0001197.}

\lref\dorey{N.~Dorey,
``An elliptic superpotential for softly broken $\cn = 4$ 
supersymmetric  Yang-Mills theory,'' hep-th/9906011,
\jhep{9907}{1999}{021}.}

\lref\dorkum{
N.~Dorey and S.~P.~Kumar,
``Softly-broken $\cn = 4$ supersymmetry in the large-$N$ limit,''
hep-th/0001103, \jhep{0002}{2000}{006}.}

\lref\swone{N.~Seiberg and E.~Witten,
``Electric - magnetic duality, monopole condensation, and confinement
in $\cn=2$ supersymmetric Yang-Mills theory,'' hep-th/9407087,
\np{426}{1994}{19}.}

\lref\swtwo{N.~Seiberg and E.~Witten,
``Monopoles, duality and chiral symmetry breaking in $\cn=2$ 
supersymmetric QCD,'' hep-th/9408099, \np{431}{1994}{484}.}

\lref\DKM{ N. Dorey, V. V. Khoze and M. P. Mattis, 
``On mass-deformed $\cn=4$
supersymmetric Yang-Mills Theory,'' hep-th/9612231,
\plb{396} {1997} {141}.}

\lref\thooft{G. 't Hooft, ``A property of electric and magnetic flux
in nonabelian gauge theories,'' \np{153}{1979}{141}.}

\lref\dvashi{G. Dvali and M. Shifman, ``Domain walls in strongly
coupled theories,'' hep-th/9612128, \plb{396}{1997}{64}.}
\myTitle{\vbox{\baselineskip12pt
\hbox{hep-th/0006008}\hbox{RUNHETC-2000-18}\hbox{SWAT-261}
\hbox{UW/PT-00-05}}}
{\vbox{
{\centerline {New Modular Invariance in the $\cn=1^*$ theory,}}
\vskip .1in
{\centerline {Operator Mixings and Supergravity Singularities}}
  }}
\centerline{Ofer Aharony$^{(a)}$\foot{oferah@physics.rutgers.edu}, Nick
Dorey$^{(b)}$\foot{n.dorey@swansea.ac.uk} and S. Prem
Kumar$^{(c)}$\foot{kumar@phys.washington.edu}} 
\vskip.1in 
\centerline{ {\it (a) Department of Physics and Astronomy, Rutgers University,
Piscataway, NJ 08855, USA}}
\centerline{ {\it (b) Department of Physics, University of Wales Swansea,
Singleton Park, Swansea, SA2 8PP, UK}}
\centerline{ {\it (c) Department of Physics, University of Washington,
Box 351560, Seattle, WA 98195-1560, USA}}
\vskip .2in

\centerline{\bf Abstract}

We discuss the mass-deformed $\cn=4$ $SU(N)$ supersymmetric Yang-Mills 
theory (also known as
the $\cn=1^*$ theory). We analyze how the correlation functions of this
theory transform under $S$-duality, and which correlation
functions depend holomorphically on the complexified 
gauge coupling $\tau$. We provide
exact modular-covariant expressions for the vacuum expectation 
values of chiral operators in the
massive vacua of the $\cn=1^*$ theory. 
We exhibit a novel modular symmetry of the chiral sector
of the theory in each vacuum, which acts on the coupling $\tt=
(p\tau+k)/q$, where $p$, $k$ and $q$ are integers which label the
different vacua. In the strong coupling limit, we
compare our results to the results of Polchinski and Strassler in the
string theory dual of this theory, and find non-trivial agreement
after operator mixings are taken into account. In particular we
find that their results are consistent with the predicted modular 
symmetry in $\tt$. Our results imply
that certain singularities found in solutions to five dimensional
gauged supergravity should not be resolvable in string theory, since 
there are no field theory vacua with corresponding vacuum expectation 
values in the large $N$ limit.

\Date{6/00}

\newsec{Introduction and summary}

The $AdS$/CFT correspondence \refs{\juan,\gkp,\wittenAdS} (see \magoo\
for a review) is
conjectured to be an exact duality between certain field theories and
certain compactifications of string/M theory. 
As originally stated the correspondence applied only to conformal field
theories, but it can easily be 
generalized also to relevant deformations of the conformal field
theories, which are realized as solutions of string/M theory with
particular boundary conditions depending on the deformation
parameter. The $AdS$/CFT correspondence relates weakly coupled field
theories to highly curved backgrounds of string/M theory, and vice
versa, a fact which makes it quite difficult to test. Most of the
tests of the correspondence so far are either qualitative in nature,
or they involve quantities that do not depend on the coupling. 
While the $AdS$/CFT
correspondence has taught us a lot about both field theory and string
theory in regions of parameter space that were previously completely
inaccessible (such as field theories with large $g_{YM}^2 N$), it has
not yet been possible to use it to learn about more traditional
theories such as gauge theories which are weakly coupled at some
energy scale, or string theory compactifications including regions of
small curvature. One of the goals of this paper will be to try to
use field theory and the $AdS$/CFT correspondence to learn about string
theory backgrounds that include regions of small curvature (where
supergravity is valid) but where supergravity seems to break down at
some singularities, and to check whether the resulting singularities
can be resolved in string theory or not.

In this paper we will study the $SU(N)$ $\cn=4$ supersymmetric
Yang-Mills (SYM) theory in four dimensions, deformed by a superpotential
which gives a mass to the adjoint scalars and to three of the four
adjoint fermions and breaks the supersymmetry to $\cn=1$; this theory
was dubbed $\cn=1^*$ in \polstr. The deformed theory has a finite (but
large for large $N$) number of possible vacuum states preserving
supersymmetry, some of which have a mass gap while others do not. 
On the string theory side of the $AdS$/CFT correspondence, 
these theories were studied in \gppz\ and in \polstr.
We
 will study this deformation by field theory methods, 
using the results
 of \refs{\dorey,\dorkum} which allow 
exact computations of vacuum expectation
values (VEVs) of various chiral operators in these theories, in each vacuum 
and for every value
of the coupling, as well as computations of
the tensions of BPS-saturated domain walls
interpolating between pairs of vacua. 
For large $N$ and strong 't Hooft coupling 
these theories are dual, by the $AdS$/CFT
correspondence, to string theory backgrounds that asymptote to
$AdS_5\times S^5$ with a large radius of curvature. We will
compare the field theory results to results found using supergravity
in this limit.

The authors of 
\polstr\ looked for solutions of
ten dimensional type IIB
supergravity (SUGRA) with possible 5-brane sources (but without any other
singularities) that would correspond to
vacua of the mass-deformed theory. They found a large number of
solutions for which the supergravity (+ 5-brane sources) approximation
could be trusted (for a large range of possible values of the
coupling), which are approximate string theory backgrounds
corresponding to many of the field theory vacua. The solutions found
in \polstr\ are isolated (up to the choice of the asymptotic string
coupling) and were identified with particular vacua of the
field theory. The authors of \polstr\ were also able to construct 
supergravity solutions corresponding to BPS domain walls interpolating
between distinct vacua. The tensions of these objects were found to
agree with the field theory predictions of \dorkum. In the following 
we will perform a similar comparison between gauge theory and SUGRA 
for the chiral condensates. In particular, using 
the solutions of \polstr\ we can compute the VEVs
of various operators at strong coupling and compare with the exact
field theory results, allowing a quantitative test of the $AdS$/CFT
correspondence at strong coupling. 
The comparison of condensates is
complicated by the fact that the symmetries allow for non-trivial
mixings among the chiral operators after the mass deformation, so
there is no unique way to define the field theory operators 
corresponding to supergravity fields. 
The resolution of this ambiguity involves 
a novel modular symmetry of the ${\cal N}=1^*$ theory which we now
discuss.  

As reviewed below, the ${\cal N}=1^*$ $SU(N)$
theory has a total of $\sum_{d|N} d$ massive vacua, labeled 
by three non-negative integers, $p$, $q$, and $k<q$, with $pq=N$.  
These vacua are permuted by $SL(2,\IZ)$
modular transformations acting on the complexified coupling constant 
\donwit\ 
$\tau=4\pi i/g^{2}_{YM}+\theta_{YM}/2\pi$. Thus, the $S$-duality 
of the underlying ${\cal N}=4$ theory is not a symmetry of the 
${\cal N}=1^*$ theory in a given vacuum, but rather 
relates the physics in distinct vacua at
different values of $\tau$. However, it turns out that a sector of
the theory 
in each vacuum is actually invariant under a different modular group
which acts on the coupling $\tilde{\tau}=(p \tau +k)/q$. As we review
below, this symmetry, which we will refer to as $\tilde{S}$-duality, 
has a simple explanation in terms of the hyperelliptic curves given by 
Donagi and Witten \donwit\ which govern the Coulomb branch of the 
corresponding theory with eight supercharges. This connection suggests
that $\tilde{S}$-duality is probably only a symmetry of the chiral sector
of the theory, which is controlled by the corresponding curve.  
We find that there is a unique definition 
for each chiral operator which transforms with definite weight under 
$\tilde{S}$-duality. In some vacua, $\tilde{S}$-duality
transformations relate two different regimes where the 
SUGRA solutions of \polstr\ are reliable. These cases 
yield non-trivial agreement between gauge theory and SUGRA. 
More generally, it is notable that in the
SUGRA approximation the solutions of \polstr\ only depend on
$\tt$ and not on $p$, $q$, $k$ and $\tau$ separately. 
This exclusive dependence on $\tt$ is itself a necessary condition 
for $\tilde{S}$-duality.       

The $AdS$/CFT dual of the ${\cal N}=1^*$ theory has 
also been studied by different methods in \gppz.
These authors looked for solutions of the 
five dimensional $\cn=8$ gauged supergravity that had
the correct asymptotic behavior to describe the mass-deformed $\cn=4$
SYM theory, and that allowed for VEVs
of some (but not all) of the
operators of the theory. They found a class of solutions with a
continuous VEV for the supergravity mode corresponding to a linear combination
of the gluino
condensate and other fields. These solutions all develop a naked
singularity at some value of 
the radial coordinate (the solutions are found by integrating the equations of
motion in the radial direction, starting from particular boundary conditions
set by the deformation parameters and the operator
VEVs). Assuming that $d=5$, $\cn=8$ supergravity is a consistent
truncation of type IIB supergravity on $AdS_5\times S^5$, the solutions of
\gppz\ may be lifted to solutions of type IIB supergravity, which
presumably also involve a naked singularity. Due to the presence of
the singularity it is not clear if these solutions correspond to
consistent string theory backgrounds (which would resolve the
singularity in some way) or not. Since the solutions of
\gppz\ feature a continuous VEV for the gluino condensate, while the
field theory has (for finite $N$) just a finite number of possible
vacua, it is clear that for finite $N$ the solutions of \gppz\ cannot
all be physical. However, since the number of vacua grows rapidly in
the large $N$ limit (as $e^{\sqrt{N}}$), it is possible for the
discrete vacua to look like a continuum in this limit, so that every
solution found in \gppz\ might be the limit of some series of allowed
vacua. If this were true, it would mean that the singularities found in \gppz\
could all 
be resolved by string theory (since we would have a consistent
string theory solution corresponding to every vacuum of the field
theory). Our field theory analysis will lead us to a different
conclusion. It seems that only one of the solutions of \gppz\
corresponds to the large $N$ limit of a series of vacua of the field
theory, while the other solutions do not seem to correspond to such a
limit, so presumably the singularities which appear in them should not
be resolvable in string theory. We will not be able to prove this
rigorously since we have field theory results only for some of the
vacua of the theory, namely the vacua with a mass gap and a small
number of massless vacua. However, we will be able to show that none
of these vacua converge to generic solutions of the type found in
\gppz, and we consider it unlikely that other series of vacua would
have a good large $N$ limit (though it is not impossible).

At first sight, one would think that there should be a correspondence
between the solutions found in \polstr\ and in \gppz. However,
generally this is not the case because the solutions of \polstr\
include 5-brane sources while those of \gppz\ do not. In many of the
vacua discussed in \polstr, there are 5-branes sitting at a radial
position of order $\sqrt{N}$ (in units of the asymptotic $AdS$ radius of
curvature), which significantly change the solutions, so there is no
sense in which these solutions of \polstr\ are related to the
solutions of \gppz. In this case it seems that instead of resolving
the singularities found by \gppz, string theory replaces them by
completely different spaces which do not resemble the solutions which
become singular. For some particular vacua, all the 5-branes of
\polstr\ sit at a finite radial position in the large $N$ limit, and
then the solutions of \polstr\ should be similar to those of \gppz,
at least at large radial positions where the effects of the 5-branes
are small (and for particular choices of parameters leading to the
VEVs analyzed in \gppz). Thus, in these cases one might say that
string theory resolves the singularities of \gppz\ by replacing them
with 5-branes. Unfortunately, as one tries to ``push'' the 5-branes to
smaller radial positions (where the solutions of \gppz\ become
singular) the approximations used in \polstr\ break down, so more work
is needed to understand exactly how the singularities are resolved by
string theory.

We will begin in section 2 by reviewing the spectrum of chiral
operators of $\cn=4$ SYM and their modular transformation properties
under the $SL(2,\IZ)$ electric-magnetic duality of the theory, which
will be useful for our analysis. In section 3 we introduce the
mass-deformed theory and discuss which correlation functions in this
theory should have a holomorphic dependence on the gauge coupling
$\tau$. In section 4 we review the field theory results of
\refs{\dorey,\dorkum} for the domain wall tensions and also obtain expressions
for operator
VEVs in different vacua, and we verify their consistency with modular
invariance. In section 5 we discuss operator mixing and the new
modular symmetry of each vacuum state described above, and show how to
define chiral operators which are covariant under this symmetry.  
In section 6 we analyze the VEVs in supergravity limits,
and show that the supergravity results of \polstr\ agree with the
field theory results for the operator VEVs if we use the 
$\tilde S$-covariant definition of the operators.
In sections 7 and 8 we look for vacua of the field theory which
could correspond to the solutions described by \gppz, and show that
some of the singularities found in \gppz\ should be resolvable in
string theory (presumably by replacing them by branes as in \polstr)
while others should not be resolvable. We also comment on the possible
relation of this result to other singular backgrounds which have
recently been discussed in the context of solutions to the
cosmological constant problem.

\newsec{Modular properties and normalizations of $\cn=4$ operators}

We begin by reviewing some basic
facts regarding the $\cn=4$ theory with $SU(N)$
gauge group in four dimensions. 
Schematically, the Lagrangian for this theory is 
\eqn\lag{{\cal L} =N{1\over {g_{YM}^2 N}} \tr\left(-{1\over 4}F^2_{\mu\nu} +
{\cal D}^\mu \phi^I {\cal D}_\mu \phi^I + [\phi^I, \phi^J]^2 + {\rm
fermions}\right) + {\theta\over{8\pi^2}}\tr(F \wedge F).}

The $\phi^I$ $(I=1,\ldots 6)$ are $SU(N)$-adjoint scalars in the $\bf 6$
representation of the global $SO(6)$ R-symmetry group.
This theory possesses superconformal invariance and is also believed to be
invariant under $SL(2,\IZ)$ transformations which  act on 
$\tau \equiv 4\pi i
/ g_{YM}^2 + \theta /  2\pi$ as $\tau \to {\tau^\prime} 
= (a\tau + b) / (c \tau + d)$. Consequently, the theory with 
parameter $\tau$
describes the same physics as the theory with parameter ${\tau^\prime}$ (on
$\IR^4$). 

The $\cn =4$ theory can also be described as an $\cn=1$ theory 
with 3 adjoint
chiral multiplets 
$\Phi_i$ and a superpotential 
proportional to $\tr(\Phi_1[\Phi_2,\Phi_3])$. The
scalar components of these superfields may be written in terms of the real 
fields $\{\phi^I\}$ as $\Phi_i=(\phi^{i}+i\phi^{i+3})/\sqrt 2$ (we use the
same symbol $\Phi_i$ for the $\cn=1$ chiral superfield and for its 
lowest component).

Following Intriligator \ken, it is convenient to define the lowest
components of the chiral primary superfields of the $\cn=4$
superconformal algebra to be 
\eqn\low{\co_p^{I_1I_2\ldots I_p}\equiv N (g_{YM}^2 N)^{-p/2}
\tr(\phi^{\{I_1}\phi^{I_2}\ldots \phi^{I_p\}}).} 
Here the $I_j$ are $SO(6)$ indices contracted to form a symmetric
traceless product of the $p$ $\bf6$'s, 
corresponding to representations of weight
$(0,p,0)$ of the $SU(4)\simeq SO(6)$ R-symmetry group. 
In the normalization of \low\ the chiral primary operators $\co_p$ are
$SL(2,\IZ)$ invariant\foot{With the kinetic 
terms normalized as in \lag, this is 
clearly true for instance in the case of the $U(1)$ theory which is free.}. 
Furthermore, given the normalization \low, in the large $N$ limit
all correlation functions of $\co_p$'s and their descendants are proportional to
$N^2$, and they also have a smooth limit as $\lambda \equiv
g_{YM}^2 N \to 0$. This is the appropriate leading behavior for correlators in the
large $N$ limit, implying that these operators in the $\cn=4$ theory can
be directly identified with type IIB supergravity fields without
introducing any additional factors of $N$.

Since the operators $\co_p$ are $SL(2,\IZ)$ invariant, the
fields $\tr(\phi^p)$ 
(omitting the $SU(4)_R$ indices for brevity) transform under
$SL(2,\IZ)$ as modular forms with modular weight\foot{$\co$
is a modular form of weight $(w,{\bar w})$ if when $\tau \to (a
\tau + b) / (c\tau + d)$, 
$\co \to (c\tau + d)^w (c{\bar \tau}+d)^{\bar w} \co$.}\
$(p/2,p/2)$ 
(recall that $\Im(\tau)$ is a modular form of weight $(-1,-1)$).
The $\cn =4$ supercharges $Q^A_\alpha$ 
and ${\bar Q}_{A\dot{\alpha}}$ can also be
thought of effectively as modular forms with weight $({1\over 4},-{1\over4})$
and $(-{1\over 4},{1\over 4})$ respectively \ken. Therefore,
the descendant fields
obtained by acting on $\co_p$ as defined in \low\ with $k$ powers of 
$Q$ and $l$ powers of ${\bar Q}$ transform as modular forms of weight
$((k-l)/4,-(k-l)/4)$.

In what follows we will be primarily interested in the VEVs for certain chiral
operators (in the $\cn=1$ sense) in the mass-deformed version of
$\cn=4$ SYM.  
The operators of interest are some of the $\cn=4$ operators $\co_2$
(with a special choice of $SU(4)_R$ indices so that 
$\co_2=\co_{2}^{(i)}\equiv\tr(\Phi_i^2)/g^2_{YM}$), 
and a particular scalar descendant of
this operator obtained by acting on  
it with two supercharges of the same chirality. 
We will schematically denote this
descendant as $Q^2 \co_{2}$, which is shorthand for the nested commutator 
$\{Q^{A}_{\alpha},[Q^B_\beta,\co_2^{(i)}]\}\epsilon_{\alpha\beta}$. 
This operator
will be discussed in more detail below. The operator
$\co_2^{(i)}$ is a component of 
$\tr(\phi^{\{I}\phi^{J\}}) / g_{YM}^2$, which is in the 
$\bf{20'}$ representation of $SU(4)_R$ and is $SL(2,\IZ)$ invariant, while
$Q^2\co_2$ is a complex operator in the $\bf 10$ representation of $SU(4)_R$ 
and is a modular form of weight $(1/2,-1/2)$. Note that
$\tr(\Phi^2_i)$ itself is a modular form of weight $(1,1)$.

Naively (as reviewed in \magoo) the action of the two
supercharges on $\co_2$ gives
rise to two types of terms. When the supercharges each act on a distinct term
in the bilinear, a symmetric product of two fermions $\psi$ of the same
chirality is obtained. The action of  
both supercharges on the same scalar gives (choosing an 
$\cn=1$ subalgebra) $F$ and $D$ terms of the form
$[\phi^I,\phi^J]$. 
Thus, classically one expects that the various components of $Q^2\co_2$ in the
$\cn=4$ theory must be of the schematic form ${\sim \tr(\psi \psi + \phi^I
[\phi^J, \phi^K]),}$  
where the indices on the fermions and scalars are contracted to give an
operator in the $\bf 10$ representation of $SU(4)_R$ (if we choose $Q$ to
transform in the $\bf{\bar 4}$ representation). 

Specifically, choosing an $\cn=1$ subalgebra and denoting the gluino by
$\psi_4$, for a particular choice of $SU(4)$ indices the action of two
supercharges $Q^A_\alpha$ on $\co_2^{(i)}$ yields 
\eqn\descone{Q^2\co_2\bigg|_{classical}\equiv {1\over {g^2_{YM}}}
Q^2\tr(\Phi_i^2)=
{1\over {g^2_{YM}}}\tr\left(\psi_4\psi_4+
2\Phi_1[\Phi_2,\Phi_3]\right).}
As indicated, this expression is valid only classically.
 The double action of the supercharges on the composite operator $\co
_2^{(i)}$ also gives rise to an anomalous term which shows up at
one loop. In $\cn=1$ supersymmetric gauge theories in general, this term is
known as the Konishi anomaly \refs{\konishione,\konishitwo}, 
and in the theory with adjoint fermions it is 
proportional to $g^2_{YM}N \tr(\psi
\psi)$ (with a  coefficient of order one). 

The existence of this term can be shown by regulating operator products
involving $Q^2\co_2$ via a convenient regularization scheme,
such as Pauli-Villars. The
finite contribution from the Pauli-Villars regulator fields then leads to the
Konishi anomaly, as in \konishione. Therefore, we find that 
$Q^2 \co_2$ is given by 
\eqn\desctwo{Q^2 \co_2={1\over{g^2_{YM}}}\tr\left(\psi_4\psi_4+
2\Phi_1 [\Phi_2, \Phi_3] + g^2_{YM}N K\right),} 
where $K$ represents the Konishi anomaly. Note the
relative factor of 2 between the first and second terms, which will be   
significant below. In the weak coupling limit
$K\sim\tr(\psi_4\psi_4)$. Although the Konishi anomaly may be 
argued to be 1-loop exact in certain cases, we do not discount 
the possibility
of further perturbative and non-perturbative contributions. 
In fact, the exact
form of the anomaly will not concern us. 
What will turn out to be important is that the same anomaly also appears in 
$\bar Q_{\cn=1}^2\tr(\bar \Phi_i \Phi_i)$,
where $Q_{\cn=1}$ and $\bar Q_{\cn=1}$ represent
the unbroken supercharges of the mass-deformed $\cn=4$ theory.

\newsec{$SL(2,\IZ)$ and holomorphy in the $\cn=1^*$ theory}

The mass-deformed $\cn=4$ theory
(henceforth referred to as the $\cn=1^*$ theory as in \polstr) 
is obtained by
introducing explicit mass terms for the 
adjoint chiral  multiplets, such that the superpotential is
\eqn\superpot{W=-{\tau\over{16\pi i}} \tr(W_\alpha^2)+
{1\over g^2_{YM}}\;\tr\left(\Phi_1[\Phi_2,\Phi_3]+m_1\Phi_1^2+m_2\Phi_2^2+m_3\Phi_3^2\right).}
The theory has ${\cal N}=2$ SUSY in the special case where one of the
three masses vanishes and the other two are equal. We will refer to
this as the ${\cal N}=2^*$ theory.   
Although no special holomorphy properties of the $\cn=4$ theory or its
mass deformation are apparent from \lag\ 
and \superpot, it can be argued in the $\cn=1$ and
$\cn=2$ languages that certain correlation functions in these
theories depend holomorphically on $\tau$ and on the mass
parameters. These holomorphy properties 
are not evident in \lag\ and \superpot\ since the
gauge kinetic term (and terms related to it by 
$\cn=1$ or $\cn=2$ supersymmetry), the kinetic
terms for the chiral multiplets and the superpotential 
are all proportional to $1/g_{YM}^2 \sim \Im(\tau)$. Thus, naively one
would not expect correlators in the theory to have holomorphic dependence on
$\tau$.  

To get a holomorphic dependence on $\tau$ we need to normalize the
fields so that $\Im(\tau)$ does not appear in the superpotential. Since
the $\cn=4$ superpotential is proportional to 
$\tr(\Phi_1 [\Phi_2,\Phi_3])/g_{YM}^2$, 
and we want to get a superpotential proportional to $N$ but
not to $1/g_{YM}^2$, we need to work in terms of rescaled fields $\tp_i \equiv
\Phi_i / \lambda^{1/3}$. The kinetic terms of the $\tp_i$ have an ugly
dependence on $g_{YM}$, but the superpotential is now
simply \eqn\super{W = \tr[-{\tau \over {16\pi i}}
W_{\alpha}^2 + N(\tp_1 [\tp_2, \tp_3] + \tm_1
\tp_1^2 + \tm_2 \tp_2^2 + \tm_3 \tp_3^2)],} 
where $\tm_i \equiv m_i / \lambda^{1/3}$ (note that due to the non-standard
normalization of the kinetic terms, the physical masses of the chiral 
superfields are
still given by $m_i$ and not by $\tm_i$).
General arguments (such as promoting $\tau$ in the superpotential to a 
chiral multiplet whose lowest component has a VEV $\tau$)
can now be used to show that 
correlation functions of the fields $\tp_i$ and
$W_{\alpha}$ should be holomorphic in $\tm$ and $\tau$.

In summary, while the natural fields 
and physical parameters to use in these theories
are $\Phi_i$ and $m_i$ respectively, the holomorphy properties become clear
when the theory is rewritten in terms 
of the fields $\tp_i$ and parameters $\tm_i$.
The modular properties of the operators in the $\cn =1^*$ theory can now be
easily deduced by examining the relationship between the two sets of fields and
parameters. 

As in \superpot, the mass deformation naturally appears with a coefficient
$1/g_{YM}^2$ like all the other terms in the $\cn=4$ Lagrangian. 
The parameter $m_i$ therefore couples (at leading order in the mass) to a
field of the form $Q^2
\tr(\Phi_i^2) / g_{YM}^2$, which is a component of 
$Q^2 \co_2$. As described above, this operator has
the appropriate normalization for comparing with supergravity and 
has modular weight
$(1/2,-1/2)$. Therefore the $m_i$ must naturally have modular weight
$(-1/2,1/2)$ to ensure modular invariance (i.e. we can assign modular
transformations with this weight to the $m_i$ in order to preserve modular
invariance after the mass deformation). The $m_i$ are related to the natural
holomorphic parameters by $\tm_i = m_i / \lambda^{1/3} \propto m_i
(\Im(\tau)/N)^{1/3}$, 
so $\tm_i$ has modular weight $(-5/6,1/6)$.

 From the above analysis,
gauge invariant operators 
of the form $u_p \simeq \tr(\Phi^p)$ with no additional dependence on $g_{YM}$
transform as modular forms of weight $(p/2,p/2)$ in the $\cn=4$ theory \ken. 
These fields are related to the 
holomorphic coordinates $\tu_p \simeq \tr(\tp^p)$ by $\tu_p =
u_p / \lambda^{p/3} \propto (\Im(\tau)/N)^{p/3} u_p$. 
Thus, the natural holomorphic fields $\tu_p$ have modular weight $(p/6,p/6)$.
Holomorphy now determines relations like $\tu \sim \tm^2 e_i(\tau)$
 for the location of the massive vacua of the mass-deformed
$SU(2)$ $\cn=4$ theory \swtwo\ in
which both sides have the same modular weight
($(1/3,1/3)=2(-5/6,1/6)+(2,0)$); these lead to similar relations 
$u \sim m^2 e_i(\tau)$ for
the more standard fields after rescaling by appropriate powers of
$\lambda$.

Before moving on to a detailed discussion of the condensates in this theory we
would like to clarify the relation of the chiral fields $u_p$ above, with
modular 
weight $(p/2,p/2)$, to the gauge-invariant parameters with modular weight
$(p,0)$ appearing in the
Donagi-Witten curves \donwit\ for the
$\cn=4$ theory with an $\cn=2$-preserving adjoint mass term.
We note that the variables $x,y$ and $t$ in the Donagi-Witten curves
$F(t,x,y)=0$ can be rescaled by powers of the masses such that the curves only
depend on $u_p/m^p$. Choosing mass parameters that transform as $(-1/2, 1/2)$
forms as above then implies that the quantities $u_p/m^p$ on which the curves
depend have weight $(p,0)$. This agrees with the conventions of
Donagi and Witten, who use modular-invariant masses and gauge-invariant fields
$u_p$ transforming as $(p,0)$ forms so that the dimensionless fields $u_p/m^p$
have weights $(p,0)$.
A related issue is that while one would expect the holomorphic fields $\tu$
and $\tm$ 
to have holomorphic modular weights, we found that they do
not. However, if we assign R-charges to the masses so that 
the mass-deformed theory has a $U(1)_R$ symmetry with charge $2/3$ for
$\tp$ and $\tm$, then the anti-holomorphic weights of all the holomorphic
fields are just $1/4$ of their R-charge. The expectation value of
any combination of $\tu_p$'s
is given by a polynomial in the masses
with the same R-charge times some function of $\tau$, and this function
will always have a purely holomorphic modular transformation as expected.

\newsec{Vacua and condensates in the $\cn=1^*$ theory}

Having identified the relation between the holomorphic operators and the 
operators with good modular transformation properties,
we can now compute VEVs for the
latter in the $\cn=1^*$ vacua using the known results for the holomorphic
operators. Up to possible operator redefinitions which will be
discussed below,
the condensates of  
the holomorphic operators $\tr(\tp_i^2)$ and $\tr (W_\alpha^2)$ 
can be determined from \dorey\ where an exact 
superpotential for the mass-deformed
$\cn=4$ theory was obtained. 
Since this superpotential is holomorphic, the results
above imply that it should be written in terms of
the parameters $\tm_i$. The effective superpotential of the
$SU(N)$ $\cn=1^*$ theory on $\IR^3\times S^1$ is of the form \dorey\ 
\eqn\tdsup{W = N \tm_1 \tm_2 \tm_3 \sum_{1 \leq a < b \leq N} {\cal
P}(X_a-X_b | \tau),}
where the $X_a$ ($a=1,\cdots,N; \sum_a X_a=0$)
are the chiral superfields arising from the gauge
field after compactifying on
a circle at a generic point on the Coulomb branch.
 From this superpotential we can read off the four dimensional
superpotential in every vacuum of the theory (corresponding to an extremum
of \tdsup). The massive vacua of the
theory are labeled \refs{\vafwit,\donwit}
by possible factorizations of $N$ as $N = pq$ and by an
integer $k=0,\cdots,q-1$, and the value of the superpotential in these
 vacua turns out to be\foot{In the case where $N$ is a prime number, this
result, as well as the corresponding formulae for the VEVs of ${\cal
N}=1^*$ chiral condensates, appears in \dorkum. The general 
result was independently derived in \polstr.}
\eqn\fdsup{W = {{N^3}\over {24}} \tm_1 \tm_2 \tm_3 [ E_2(\tau) -
{p\over q} E_2( {p\over q} \tau + {k\over q})].}
The second Eisenstein series $E_2(\tau)$ \koblitz\ is the 
holomorphic function which is closest to being a modular form of
weight two; its modular transformation properties are
$E_2(\tau+1) = E_2(\tau)$ and $E_2(-1/\tau) =
\tau^2 E_2(\tau) - 6i\tau/\pi$. 

It should be noted that within field theory there is actually an ambiguity in
the superpotentials \tdsup\ and 
\fdsup, corresponding to a possible additive holomorphic contribution \dorkum\  
$A(\tau, N)$, with a weak coupling expansion of the form
$\alpha_0+\sum_k\alpha_k e^{2\pi ik\tau}$. 
As the shift is independent of the chiral superfields
$X_{a}$, it is the same in each vacuum of the theory. 
In particular, the ambiguity corresponds to the addition of an 
arbitrary holomorphic function of $\tau$, which is independent of $p$,
$q$ and $k$, to the superpotential \fdsup. However, it is worth noting that 
any such shift will spoil the modular transformation properties of 
$W$ (apriori there is no reason for $W$ to have nice modular
transformation properties, but as discussed below $\del W/\del {\tilde
m}_i$ is naturally a modular form, and in the $\cn=1^*$ theory $\del W/\del
{\tilde m}_i = W / {\tilde m}_i$). 
Indeed, to preserve the modular properties of \fdsup, the
function $A$ would need to be a holomorphic modular form of 
weight two. Here we are assuming that the superpotential does not 
have any unphysical singularities in the interior 
of the fundamental domain of $SL(2,\IZ)$. 
As no such forms exist, we see that \fdsup\ is the unique modular 
definition of the holomorphic superpotential. The ambiguity can also 
be understood from the point of view of the ${\cal N}=2^*$ theory,
where it corresponds to a freedom in defining different coordinates on
the Coulomb branch.  In fact, for gauge group $SU(2)$, 
the additive shift described above corresponds
 precisely 
to the mismatch between the modular covariant Seiberg-Witten
coordinate \refs{\swtwo,\DKM} and the physical VEV 
$\vev{\tr(\Phi_i^2)}$. From this point of view it is also obvious that
the ambiguity is vacuum independent. As we discuss extensively below, 
this reflects a more general phenomenon: 
operators which are uniquely 
defined in the ${\cal N}=4$ theory can mix with mass-dependent 
coefficients once the mass deformation is turned on.   

The total number of massive vacua of the $\cn=1^*$ theory is 
given by the sum over divisors of $N$, $\sum_{d|N}d$. In fact, this theory
realizes every possible massive phase seen in 't Hooft's abstract
classification of phases of $SU(N)$ gauge theories with a $\IZ_N$
symmetry \thooft.
The standard ``Higgs'' vacuum in which the gauge group is completely
broken classically arises for $p=N,q=1$, while the
``confining'' vacua (for which classically $\Phi_i=0$),
which are related to the vacua of the $\cn=1$ pure SYM theory in an
appropriate weak-coupling limit, arise for $p=1,q=N$. The ``Higgs''
vacuum corresponds to a condensation of electric charges and
confinement of magnetic charges, while the ``confining'' vacua
correspond to a condensation of magnetic (or dyonic) charges and a
confinement of electric charges.
For prime $N$ there are no
additional massive vacua. The weak coupling expansion of $E_2$ implies that
vacua corresponding to integer values of $p/q$ have a weak
coupling expansion which can be interpreted as arising purely from
instanton corrections; in other vacua additional nonperturbative
effects, which may perhaps
be associated with fractional instantons or merons, appear. 
Modular transformations act on the vacua by a non-trivial permutation,
which follows from identifying the electric and/or magnetic charge of the
particles which condense in each vacuum.

 From \fdsup\ we can obtain the tensions of BPS-saturated
domain walls between two 
different vacua, by computing the absolute values of superpotential differences
between the vacua \dvashi.
The superpotential difference between generic massive vacua is
\eqn\deltatwo{\Delta W_{1,2} = {N^2 \over 24} \tm_1 \tm_2 \tm_3
[p_1^2E_2({p_1\over q_1}\tau+{k_1\over q_1}) - 
p_2^2E_2({p_2\over q_2}\tau+{k_2\over q_2})].}
One can check that this is a modular form of weight
\eqn\deg{3(-5/6,1/6)+(2,0)=(-1/2,1/2)} 
up to permutations of the vacua, so the
domain wall tensions $T=|\Delta W|$ are modular invariant up to
permutations, as they should be. In terms of the standard mass parameters
the coefficient of these
expressions involves $m_1 m_2 m_3 / \lambda =
\Im(\tau) m_1 m_2 m_3/4\pi N$ 
instead of $\tm_1
\tm_2 \tm_3$; the $m_i$'s are the physical masses, 
and
they are also the natural deformation parameters in SUGRA as described above.

The superpotential \fdsup\ also determines the values of the
condensates of the chiral operators $\tu_2^{(i)}\equiv
\tr(\tp_i^2)$ and $S\equiv\tr (W_\alpha^2)$ in the various $\cn=1^*$
vacua. This is a consequence of the usual
identifications 
\eqn\cond{\vev{\tu_2^{(i)}}={1\over N}{{\del W}\over{\del \tm_i}};
\quad \quad \quad \vev{S}=-16\pi i{{\del W}\over{\del \tau}}.} 
In the normalization of \super, $S$ is
also a chiral superfield of the $\cn=1^*$ theory, and its
correlation functions should be holomorphic in $\tau$ and in the
$\tm_i$. Assuming equal masses\foot{It is easy to generalize our results
to arbitrary masses just by using the global symmetries, which determine
for example that $\vev{\tu_2^{(1)}}\propto m_2 m_3$ and $\vev{S} \propto 
m_1 m_2 m_3$.} and using $\tm_i=m_i/\lambda^{1/3}$, 
$\tu_2=u_2/\lambda^{2/3}$, we see that  
\eqn\vevutwo{ 
\vev{u_2} ={{N^2} \over {24}} m^2
(E_2(\tau) - {p\over q} E_2({p\over q}\tau+{k\over q})),}
where we have set $A(\tau,N)=0$, thus
providing a modular covariant definition of the holomorphic condensate. 
Via \low\ this immediately leads to a modular invariant result 
 for the VEV of $\co_2$,
\eqn\vevcotwo{
{\vev{\co_2}} = {N^2 \over {96\pi}} \Im(\tau) m^2 
[E_2(\tau) - {p\over q} E_2({p\over q}\tau + {k\over q})].}
Up to permutations between the vacua of the $\cn=1^*$ theory 
this operator
 is invariant under $SL(2,\IZ)$, as expected. 

 From the definition \desctwo, the operator  
$g^2_{YM}Q^2\co_2$ of the ${\cal N}=4$ theory\foot{Note 
that $Q$ here denotes the 
supercharges of the original $\cn=4$ theory and not 
the $\cn=1$ supercharge 
 which is preserved 
after the mass deformation.} is given by
$\tr(\psi_4\psi_4)+2\tr(\Phi_1[\Phi_2,\Phi_3])+g^2_{YM}N K$, the last term
representing the anomalous contribution. The first term 
$\tr(\psi_4\psi_4)$ is the gluino condensate, which is the lowest
component of the ${\cal N}=1$ chiral superfield $S=\tr (W_\alpha^2)$. 
The VEV of this operator is holomorphic in $\tau$ and
can be obtained from the exact superpotential \fdsup\ using the second
relation in \cond .

%
%

 Before the mass deformation
the sum of the
last two terms is $Q_{\cn=1}$-exact and, therefore, cannot get a VEV. 
However, this is no longer
true in the mass deformed theory.
In fact, the VEV of these terms can be determined by
considering the operator
${\bar Q}^2_{\cn=1}\tr(\bar\Phi_1\Phi_1)$, 
which is classically proportional to
$\tr(F_1^*\Phi_1)$. Quantum mechanically it is well-known \refs{\konishione,
\konishitwo} that this is accompanied by the
anomalous piece $-g^2_{YM}NK$. The Q-exactness of this combination  
leads to a relation between the F-terms and the anomaly, of the form
\eqn\konishi{2\vev{\tr (\Phi_1F_1^*)}=
-4\vev{m_1\tr(\Phi_1^2)}-2\vev{\tr(\Phi_1[\Phi_2,\Phi_3])}
=g^2_{YM}N \vev{K}.}
This relates the VEV of $\tr(\Phi_1^2)$ to the VEV of 
$2\tr(\Phi_1[\Phi_2,\Phi_3])+g^2_{YM}N K$ which is precisely
the combination appearing in $Q^2\co_2$. Thus, we find
\eqn\qsqoqftA{{\vev{Q^2\co_2}}
={1\over g^2_{YM}}(\vev{\tr(\psi_4\psi_4)}-4\vev{m_1\tr(\Phi_1^2)})
=-{16\pi i\over g^2_{YM}}\left({{\del
W}\over{\del\tau}}-{i\over\Im(\tau)}W\right).}
This is simply $-4i\Im(\tau)\tm^3$ multiplied by the modular covariant
derivative\foot{The 
modular covariant derivative ${{\cal D}\over{\cal D}\tau}$ of a $(w,\bar
w)$ form $\co$ is given by $d{\co}/d\tau-iw\co/2\Im(\tau)$,
which in turn yields a modular form of weight $(w+2,\bar w)$.}
of the form $W/{\tm^3}$ of weight $(2,0)$. 
The expression \qsqoqftA\ has the correct modular weight 
$(1/2,-1/2)$, thus providing a modular covariant (up to permutations)
result for the VEV of $Q^2\co_2$
\eqn\qsqoqft{\eqalign
{&
\vev{Q^2\co_2}= -4i \Im(\tau)\tm^3{{\cal
D}\over{{\cal D}\tau}} 
\left({W\over \tm^3}\right)=\cr
&={N^2\over {24\pi i}}(\Im(\tau))^2m^3\left[
E_2^\prime(\tau)
-{p^2\over q^2}E_2^\prime\left({p\over q}\tau+{k\over q}\right)
-{i\over\Im(\tau)}\left\{E_2(\tau)
-{p\over q}E_2\left({p\over q}\tau+{k\over q}\right)\right\}\right].\cr}}
The identification of the SUSY charge $Q^2$ with the modular covariant
derivative appears to be a general result. It is familiar in the
context of the ${\cal R}^{4}$ term in the type IIB effective action (either
on $\IR^{10}$ \grgut\ or $AdS_{5}\times S^5$ \bagr), and it is inherited by
correlation functions of chiral primary operators in the
${\cal N}=4$ theory 
via the $AdS$/CFT correspondence. 

\newsec{Operator mixing and a new modular symmetry} 

Before proceeding, we should reconsider the additive ambiguity 
which appears in the definition of the operators discussed in the 
previous section. 
Since the ambiguity is vacuum
independent, it is convenient to describe it in terms of a 
mixing between the operator $\co_{2}$, as defined above, and 
$m^{2} {\cal I}$ where ${\cal I}$ is the identity operator. 
In particular, we may redefine $\co_2$ according to the operator 
equation \eqn\otwost{\co_{2}\rightarrow
\tilde{\co_{2}}={\co_2}+m^2\Im(\tau)f^{(2)}(\tau,\bar{\tau}){\cal I},}
where $f^{(2)}(\tau,\bar{\tau})$ is an arbitrary function of 
$\tau$ and $\bar{\tau}$ reflecting the fact that, apriori, 
$\tilde{\co_{2}}$ does not have any special holomorphy properties. 
Once again, it is convenient to restrict our attention to cases 
where $\tilde{\co_{2}}$ transforms with definite modular weight. 
This requires that $f^{(2)}(\tau,\bar{\tau})$ has modular weight 
$(2,0)$. However, unlike the holomorphic case considered above this 
is not prohibitive: there are an infinite number of possible 
choices of $f^{(2)}$ which preserve the modular properties of $\co_{2}$. 

It turns out that 
exactly one choice of $f^{(2)}$ has a very special property which
we now discuss. Specifically, if we set, 
\eqn\ftwo{f^{(2)}(\tau,\bar{\tau}) = -{N^2 \over {96\pi}}
\left( E_2(\tau)-{3 \over {\pi \Im(\tau)}} \right),}
we find 
\eqn\oper{ \vev{\tilde{\co_2}} = N^2 m^2 
\left[ {1\over {32\pi^2}} - {{\Im(\tau)}\over{96\pi}} 
{p\over q} E_2({p\over q} \tau + {k\over q}) \right].}
First, we can see that that the expression \oper\ does not
depend separately on $\tau$, $p$, $q$ and $k$, but just on the
combination $\tt \equiv (p\tau+k)/q$. Moreover, we have
\eqn\expl{
\vev{\tilde\co_2} = N^2 m^2 \left[ {1\over {32\pi^2}} - {{\Im(\tt)}\over
{96\pi}} E_2(\tt) \right],}  
which shows that, in each vacuum, $\tilde{\co_{2}}$ is actually a modular
form in the variable $\tt$ with weight\foot{This modular weight in 
$\tt$ differs from the corresponding weight under ordinary $S$-duality
transformations in $\tau$  simply because we have 
chosen to assign zero weight to the masses under the former symmetry,
while they have non-trivial weights under the latter.} $(+1,-1)$.     

A similar discussion applies to $Q^{2}
\co_{2}$. As this operator has mass dimension three, 
it can mix with $m \co_{2}$ and
$m^3 {\cal I}$ : 
\eqn\qsqotwo
{{Q^2 \tilde{\co_2}}={Q^2\co_2}+m\co_2
\Im(\tau)g^{(2)}(\tau,\bar\tau)+m^3\Im(\tau)^2 h^{(4)}(\tau,\bar\tau)
{\cal I}.}
To preserve covariance under $S$-duality transformations, 
$g^{(2)}$ and $h^{(4)}$ must be (non-holomorphic) 
modular forms of weight $(2,0)$ and $(4,0)$,
respectively. Note that this mixing (as well as \otwost) is
consistent with the $SU(4)$ R-symmetry of the $\cn=4$ theory.

As in the case of $\co_{2}$ there is a unique definition of this 
operator which is modular in $\tt$. This definition uses
$g^{(2)}=0$ and 
\eqn\goodmixings{\eqalign{
h^{(4)}(\tau,\bar\tau) 
&= -4i{{{\cal D}f^{(2)}}\over {{\cal D}\tau}} = 
 -{N^2 \over {24\pi i}} \left( E_2^{\prime}(\tau) - {iE_2(\tau) \over 
\Im(\tau)} - 
{3 \over {2\pi i (\Im(\tau))^2}} \right), \cr}}
giving
\eqn\mixedvevs{\vev{Q^2 \tilde{\co_2}} = 
N^2 m^3 \left[  -{1\over {32\pi^2}} +
{\Im(\tau)\over {48\pi}}{p\over q}
E_2({p\over q} \tau + {k\over q}) -
{(\Im(\tau))^2\over {48\pi i}}{p^2 \over q^2} 
E_2^\prime({p\over q}\tau + {k\over q}) \right].}
Rewriting this explicitly in terms of $\tt$, using the
relation $E_2^\prime(\tau) = (\pi i / 6) (E_2^2(\tau) - E_4(\tau))$,
we have
\eqn\nmixedvevs{\eqalign{
\vev{Q^2 \tilde{\co_2}} &= -{{32\pi^2}\over {N^2m}} \vev{\tilde{\co_2}}^2 +
{{N^2 m^3}\over {288}} (\Im(\tt))^2 E_4(\tt).
\cr}}
This definition is also consistent with the identification of the
supercharge $Q^{2}$ with $\Im(\tau) {\cal D}_{\tau}=
\Im(\tt) {\cal D}_{\tt}$. 
 From now on, $\tilde{\co_{2}}$ and $Q^2 \tilde{\co_{2}}$ will denote 
the specific expressions \expl\ and \nmixedvevs, which are the unique 
definitions of these operators which transform with definite weight
under modular transformations in $\tt$. We will now argue that
this corresponds to an interesting new symmetry of the theory 
which has not been noted previously. 

To understand the origin of this symmetry it is useful to start by
considering the corresponding theory with eight supercharges, denoted 
above by ${\cal N}=2^*$. The low-energy effective Lagrangian for this 
theory on its Coulomb branch is determined by the hyper-elliptic curve
given by Donagi and Witten \donwit . For gauge group $SU(N)$ the curve can be
thought of as a branched $N$-fold cover of the standard torus, 
$T(\tau)$, the latter being specified by the complex equation 
$y^{2}=(x-e_{1}(\tau))(x-e_{2}(\tau))(x-e_{3}(\tau))$. 
The Coulomb branch contains
singular submanifolds where some of the cycles of the curve degenerate
and one or more BPS states become
massless. Softly breaking ${\cal N}=2$ SUSY down to an ${\cal N}=1$
subalgebra by introducing a mass for the scalar field in the 
${\cal N}=2$ vector multiplet lifts the Coulomb branch except at
isolated singular points. More precisely, the massive vacua 
of the ${\cal N}=1^*$ theory correspond to points in moduli space 
where a maximal number of cycles degenerate. In the more familiar 
case of ${\cal N}=2$ SUSY Yang-Mills, the relevant singular curve for
each ${\cal N}=1$ vacuum is a sphere. In the present case, as
explained in \donwit, the maximally degenerate curve is 
an unbranched $N$-fold cover of the standard torus 
$T(\tau)$. Such $N$-fold covers are classified by three non-negative 
integers $p$, $q$ and $k<q$, with $pq=N$, and they are themselves complex
tori $T(\tt)$ with $\tt=(p\tau+k)/q$ as above. As usual, the low-energy
effective Lagrangian depends only on the complex structure of the
curve. It must therefore be invariant under modular transformations 
acting on $\tt$. As the chiral condensates of the ${\cal N}=1^*$
theory are also determined purely from the complex structure of the
curve, they must also exhibit this symmetry. In the following we will
refer to this symmetry as $\tilde{S}$-duality. 

Several features of ${\tilde S}$-duality are noteworthy. As we discuss
below, it is
almost certainly not an exact symmetry of the ${\cal N}=1^{*}$ theory, but 
only 
of the chiral sector of the theory which is controlled by the
Donagi-Witten curves. Even this point requires further qualification 
because the chiral sector usually denotes 
the ring of chiral
operators of an ${\cal N}=1$ theory, like those discussed in section 4, 
whose VEVs depend holomorphically on $\tau$. Indeed one usually 
refers to this as the holomorphic sector of the theory. 
The problem is illustrated by the holomorphic 
formula derived above,     
\eqn\vevutwob{ 
\vev{u_2} ={{N^2} \over {24}} m^2
(E_2(\tau) - {p\over q} E_2({p\over q}\tau+{k\over q})).}
This formula is not modular in $\tt$ for two reasons. First, 
it depends on $\tau$ as well as on $\tt$. This can easily be fixed 
by a holomorphic vacuum-independent redefinition which simply 
subtracts off the first term. More importantly, \vevutwob\ is not modular
in $\tt$ because of the anomalous transformation law of $E_{2}$
mentioned above. This is harder to rectify. In fact, the only way that 
$\tilde{S}$-duality can be restored is by a {\it non-holomorphic} 
additive redefinition, which effectively replaces $E_{2}(\tt)$ by 
$\hat{E}_{2}(\tt)=E_{2}(\tt)-3/\pi \Im(\tt)$ (and yields only a
{\it vacuum-independent} shift of $\vev{u_2}$). Thus, the new symmetry is 
only manifest after choosing a very particular non-holomorphic mixing 
of the chiral operators, as we did above. 
This feature is very reminiscent of the
holomorphic anomaly discussed in \holo . 

Unlike conventional $S$-duality which generically permutes the vacua, the new
modular symmetry is a symmetry of each vacuum. In vacuum states which are 
invariant under ordinary $S$-duality 
(this is the case when $p/q=1$ and $k=0$), the two dualities coincide 
(up to the fact that we assigned different modular weights to the
masses). One particularly
interesting case is that of the confining vacuum with $p=1$, $q=N$
and $k=0$. In terms of the 't Hooft coupling $\lambda =
g_{YM}^{2}N$, we have (for zero theta angle)
$\tt=4\pi i/\lambda$. Thus, in this vacuum, one generator of 
$\tilde{S}$-duality is an inversion of the 't Hooft coupling: 
$\lambda\rightarrow (4\pi)^2/\lambda$. As the 't Hooft coupling 
corresponds to the radius of the geometry in units of 
$\sqrt{\alpha'}$, this transformation corresponds to a novel kind 
of T-duality or mirror symmetry of the IIB background (although we 
emphasize again 
this only applies to the chiral sector of the theory).   
As noted in \dorkum, this relates the regime
of large $\lambda$, where the IIB side of the $AdS$/CFT duality is
tractable, to
the physically interesting regime of small $\lambda$, where the field
theory becomes weakly coupled at short distances.
Note that the ${\cal N}=1^*$ theory flows to the ${\cal N}=4$ SYM theory
in the UV. This theory 
is weakly interacting at small $\lambda$, but is believed to have  
quantitatively different properties at large $\lambda$. 
This makes it clear that the $\tilde{S}$-duality is not a symmetry 
of the ${\cal N}=1^*$ or ${\cal N}=2^*$ theory at all length scales. 
This is consistent with the above discussion, which suggests that 
it is a symmetry of the 
${\cal N}=2^*$ theory in the IR, or of a particular chiral sector of
the $\cn=1^*$ theory.  

\newsec{Comparison with the type IIB string theory dual}

The $AdS$/CFT correspondence provides an 
unambiguous relationship between the $\cn=4$ field theory chiral operators
and type IIB supergravity
fields. However, as we saw in the previous section, 
once we turn on the mass deformations, the definitions of the
operators on the field theory side becomes ambiguous due to
the possibility of non-trivial operator mixing. We expect a 
similar ambiguity to occur on the type IIB side. 
Importantly, a linear redefinition of the 
fields appearing in the SUGRA equations of motion
does not (by definition)
refer to any specific background. This reflects the fact 
(discussed above) that operator mixing leads to a vacuum
independent redefinition of the condensates. With this in mind, the 
main prediction of the field theory analysis in the previous 
section is that there exists a (unique) definition of the 
operators $\co_{2}$ and $Q^{2} \co_{2}$ such that they transform with 
definite modular weights under $\tilde{S}$-duality
transformations in each vacuum. In this section we will compare
this prediction against the SUGRA results of Polchinski and
Strassler \polstr . As the type IIB results are only available in certain
regimes of
parameter space we certainly will not be able to provide a direct 
demonstration that $\tilde{S}$-duality is present (in the same sense
as in the field theory discussion above)
on the string theory side of the correspondence. 
However, we will be able to perform several
non-trivial checks of this hypothesis.

According to the results of 
\polstr, many of the vacua of the $\cn=1^*$ theory can be
approximately
described in type IIB string theory via supergravity on asymptotically
$AdS_5\times S^5$  spacetimes with  sets of 5-branes arranged at
various $AdS$ radii. The mass perturbation appears as a non-normalizable
3-form field. The 5-brane sources with world volume $\IR^4\times S^2$ 
are wrapped at various angles around equators of the $S^5$ -- a
configuration that is rendered  dynamically stable by the D3-brane 
charges of the
5-branes. Massive $\cn=1^*$ vacua with $N=pq$ may be described by supergravity
solutions corresponding to $q$ D5-branes each carrying $p$ units of D3-brane
charge when $1/N\ll g^2_{YM}\ll
N/q^2$, or equivalently $1\ll \Im(p\tau/q) \ll p^2$, 
while they are appropriately described by $p$ NS5-branes each
carrying $q$ units of D3-brane charge when 
$1\ll \Im(-q/p\tau) \ll q^2$.

The SUGRA results for the chiral condensates, which we denote as 
${\vev{\co_2}}_{SG}$ and ${\vev{Q^2\co_2}}_{SG}$, can
be read off from the solutions for the metric and the 3-form $G_3$ 
(respectively)
in \polstr\ .
We note that the normalizations of \polstr\ differ by factors of $g_s$
from the SUGRA fields corresponding to $\co_2$ and $Q^2 \co_2$ 
whose kinetic
terms are of order
one, and that we need to express the results of \polstr\ in terms 
of the $AdS$
curvature radius (rather than the string scale) to match with our
normalizations. Translating the results of \polstr\ to coincide with our
normalizations, we find in the $N=pq$ vacuum with $k=0$
\eqn\polstrvev{\eqalign{
&{\vev{\co_2}}_{SG} \propto N^3 m^2 {\Im(\tau)\over {q^2}},\quad
{\vev{Q^2\co_2}}_{SG}\propto N^3 m^3 {\Im(\tau)\over {q^2}},\quad\quad
{\rm for\ } 1\ll \Im(p\tau/q)\ll p^2;\cr
&{\vev{\co_2}}_{SG} \propto N^3 m^2 {\Im(\tau)\over {p^2\tau^2}},\quad
{\vev{Q^2\co_2}}_{SG}\propto N^3 m^3 {\Im(\tau){\bar \tau}\over 
{p^2\tau^3}},\qquad
{\rm for\ } 1\ll \Im(-q/p\tau)\ll q^2.\cr}}
Note that the results in the second line differ 
by a phase from the result presented
in \polstr, but we believe that the phase in \polstrvev\ is the
correct one. 

These expressions depend on $\tau$, $p$ and $q$ 
only in the combination $\tt$. The dependence on $\tt$ alone is in fact a
property not just of these 
VEVs but of the full supergravity solutions of \polstr\ (though it does
not seem to be 
a property of the full string theory solutions when string loop
and $\alpha'$ corrections are taken into account). Whenever two
configurations with the same $\tt$ both have good descriptions in
terms of branes in \polstr, the corresponding supergravity solutions
are the same (including having branes at the same place,
which produce the same
supergravity fields around them). Clearly, the fact that the
SUGRA condensates depend exclusively on $\tt$ is in line with
the field theory predictions of the previous section. In particular,
exclusive dependence on $\tt$ is clearly a prerequisite for modular 
covariance in this parameter. Strictly 
speaking, the field theory prediction required only that the 
{\it exact} expressions 
on the string theory side should depend exclusively on $\tt$ 
(modulo field redefinitions). However, 
in the regime where \polstrvev\ is valid supergravity is a good
approximation and therefore the supergravity results should themselves
depend exclusively on $\tt$, as we indeed find from \polstrvev.

Another non-trivial check comes from noting that, at least 
for the vacua in which both $p$ and $q$ grow with positive powers of 
$N$ (subject to $pq=N$), there is an overlap between the range of
values of $\tt$ for which the first line is valid and the range of
values of $\tt$ for which the second line is valid after an
$\tilde{S}$-duality transformation $\tt \rightarrow -1/\tt$. 
It is easy to check that the expressions appearing 
in the first and second lines are precisely related by such a 
transformation, using the
modular weights $(+1,-1)$ and $(+2,-2)$ for $\co_{2}$ and 
$Q^{2} \co_{2}$ under $\tilde{S}$-duality as in the previous section.  

Actually, the agreement between the string theory and field theory
results is much stronger than this. To make a more detailed 
comparison, we will use both the weak coupling
expansion of $E_2(\tt)$,
\eqn\weak{E_2(\tt) = 1 - 24\sum_{k=1}^{\infty} {{ke^{2\pi ik\tt}}
\over {1 - e^{2\pi ik\tt}}},}
and the ``strong coupling'' expansion for 
large $\Im(-1/\tt)$, of the
form
\eqn\strong{E_2(\tt) \simeq {1\over {\tt^2}} + {6i\over \pi\tt}+
{\cal O}(e^{-2i\pi/\tt}).}
Using these identities we can see that throughout their domain of
validity, the SUGRA expressions for
the VEVs of both operators agree precisely with the 
unique $\tilde{S}$-covariant field theory expressions derived in the 
previous sections. Thus, we have 
$\vev{\co_{2}}_{SG} ={\vev{\tilde{O}_{2}}}$ and 
$\vev{Q^{2} \co_{2}}_{SG} ={\vev {Q^{2} \tilde{O}_{2}}}$ whenever 
the string theory expressions are valid. As discussed above, the field
theory results only require that these equalities should hold modulo field 
redefinitions or operator mixings. It would be interesting to understand 
the origin of the stronger result we have found. Note that the
supergravity results of \polstr\ 
generally do not agree with the naive (unmixed) VEVs that
we computed in section 4.

So far we have only discussed the results on the string theory side 
in the supergravity regime. The field theory predictions imply that, 
with suitable background independent field redefinitions, the 
{\it exact} string theory results for the condensates of 
$\co_{2}$ and $Q^{2} \co_{2}$ should agree with the field theory 
expressions \oper\ and 
\mixedvevs. Of course, it is not possible to test this directly, 
due to our ignorance about string theory in backgrounds with non-zero 
RR flux. However, certain qualitative observations can be made in
support of this proposition. In particular, in each vacuum, the field
theory results for large values of $\Im(\tt)$ have an expansion 
in powers of $\exp(2\pi i \tt)$ by virtue of \weak, while for large 
values of $\Im(-1/\tt)$ we have the $\tilde{S}$-dual expansion in powers 
of $\exp(-2 \pi i/\tt)$. It is natural to try to identify these 
exponentially suppressed terms with the contributions of string theory 
instantons on the type IIB side. Such an identification was given in 
\polstr\ and we will expand on it somewhat in the following. 

At large $\Im(\tt)$,  
the corresponding weakly-coupled type IIB background includes 
$q$ D5 branes with world volume $S^2 \times \IR^4$, where the radius of the
sphere is proportional to $p$. By a slight extension of the results
given in \polstr, one may show that the action of a single 
D-string worldsheet wrapped on the $S^2$ is precisely $2\pi i \tt$. 
Provided it preserves half of the unbroken supersymmetry (i.e. two
supercharges), this instanton configuration will contribute terms of
order $\exp(2\pi i \tt)$ to 
the lowest terms in the derivative expansion of the type IIB effective
action, which determine the large-distance behavior of the SUGRA 
fields and thereby determine the condensates. The field theory results
suggest that multiple wrappings of the sphere should also contribute. 
The fact that such configurations exist is plausibly related to the 
fact that, at least on $\IR^{10}$, 
any number of D1 branes and D5 branes can form a stable 
bound state at threshold which saturates the BPS bound and therefore 
preserves half the supersymmetry. This state then contributes as an
instanton after compactification on $S^{2}$. As there are no analogous
bound states of $(m,n)$-strings (with $(m,n)\neq (0,1)$) with D5 
branes, this also suggests an explanation of why only configurations 
involving wrapped D-strings contribute. 

In the case where $\Im(\tt) << 1$ with $k=0$, the type
IIB background includes 
$p$ NS5 branes wrapped on $\IR^{4}\times S^{2}$ with radius 
proportional to $q$. By similar arguments, wrapping the 
type IIB string worldsheet on $S^2$ yields an instanton with 
action $2\pi i/\tt$ which contributes terms of order 
$\exp(-2\pi i/\tt)$ to the condensates as expected. The corresponding 
statements about the existence of bound states of a fundamental string 
with an NS5-brane are related by $S$-duality to the $D1/D5$ case discussed 
above. Finally, in vacua with $k>0$, the instanton expansion comes
from $(p,k)$-strings wrapped on the $S^2$ factor of a $(p,k)$ 
fivebrane world volume. 

Of course, there are many other possible sources of corrections 
to the classical supergravity results. For example, we expect 
the D-instantons of the type IIB theory to contribute terms of order 
$\exp(2\pi i\tau)$ in the weak coupling limit 
$\Im(\tau) \rightarrow \infty$. Indeed, such terms are present in the
naive, unmixed, field theory expressions for the condensates, 
\vevcotwo\ and \qsqoqft. Also, there is no obvious reason why 
perturbative corrections, both in the string coupling and in 
$\alpha'$, cannot contribute. However, the prediction is not that 
these contributions are absent but rather that they can be removed 
by a background independent redefinition of the SUGRA (string theory)
fields. In the
case of D-instantons, the fact that their action is independent of the 
integers $p$, $q$ and $k$ which characterize the vacuum means that this
appears to be consistent. In fact, the operator mixing on the field
theory side has precisely the effect of removing these terms. 
In contrast, the string instantons discussed 
above have an action which depends on $p$, $q$ and $k$ and thus could
not be removed by any such redefinition. On the
field theory side, the absence of perturbative corrections in our
formulae ultimately comes from the existence of a holomorphic 
superpotential. On the string theory side we would also like to
understand why perturbative corrections in the string coupling are 
absent modulo field redefinitions. Our results suggest that 
there are quantities which are `almost holomorphic' on the type IIB side 
but suffer from a mild holomorphic anomaly which restores 
$\tilde{S}$-duality, as in \holo.

Finally, we note that the appearance of $\tilde{S}$-duality on the 
type IIB side of the correspondence is somewhat mysterious 
as there is no torus apparent in the
geometry. On the other hand, the ${\cal N}=2^*$ theory can be realized
on a network of intersecting branes in type IIA string theory \witm . 
After lifting to M-theory, the theory is realized as a
compactification of the six-dimensional $(2,0)$ superconformal theory 
which lives on the M5 brane. The compactification manifold is
precisely the relevant Donagi-Witten curve. At the singular points in
the moduli space this is a torus with complex structure 
$\tilde{\tau}$. Hence, $\tilde{S}$-duality has a natural geometrical 
realization in this construction. This is very similar 
to the T-dual realization of type 
IIB $S$-duality in M-theory on $T^{2}\times
\IR^{9}$. Presumably, soft breaking 
to the ${\cal N}=1^*$ theory could be accomplished along the lines of 
\refs{\mqcd,\mqcdoog}. 
The relation between the resulting type IIA/M theory set-up and the 
type IIB construction of the same theory discussed above is not obvious. 
As above, this is very suggestive of T-duality, 
but it would be interesting to make this more precise.

\newsec{Large-$N$ limits of Higgs and confining vacua}

In the rest of this paper we will attempt to ascertain which vacua of the
$\cn=1^*$ 
field theory could be described by supergravity solutions without 5-brane
sources, of the type analyzed in \gppz. For such vacua,
in the large $N$ limit the dual string theory solution should converge
to a fixed metric in some finite region near the boundary (say, for
the radial coordinate bigger than a fixed number times the $AdS$
radius). There are two related ways to analyze this question. 
One approach is to 
look at the scaling of the VEVs of chiral operators in a large $N$
limit (there are two such limits as we discuss below).  The existence of a
fixed 
SUGRA description in this limit would mean 
that 
correlation functions and VEVs should scale as $N^2$  
since the classical supergravity action scales as $N^2$ in
our normalization. In particular we would expect $\vev{\co_2}_{SG}$ and
$\vev{Q^2 \co_2}_{SG}$ to scale this way. An alternative approach is to 
look at the solutions of \polstr\ for the string theory duals of
$\cn=1^*$ vacua and ask when they converge to a fixed supergravity
background in the large $N$ limit. Since the positions of the 5-branes in these
solutions are linked to the VEVs of chiral operators, the two approaches are
in fact related.

There are two different types of large $N$ limits that could
correspond 
to a SUGRA 
background. 
One is the large $N$ limit with $\lambda$ fixed and large,
which corresponds to a supergravity
background in which one can have both a string perturbation expansion
and an $\alpha'$ expansion, corresponding to
first performing an expansion in $1/N$ (for
constant $\lambda$) and then in $1/\sqrt{\lambda}$. The other is a
large $N$ limit with fixed coupling, in which one still has a
$1/N$ expansion. 
In both cases we expect all correlation functions to scale as $N^2$.

Let us start by looking at the standard Higgs and confining vacua of
the $\cn=1^*$ theory. In these vacua, 
the position of the 5-branes in the solution of \polstr\ (in $AdS$
units) scales as $\sqrt{N}$ in the large $N$ limit with fixed coupling,
so the gravitational background is not fixed in this limit. In the
large $N$ limit with fixed (and large) $\lambda$, the position of the
D5-branes in the Higgs vacuum scales as $N/\sqrt{\lambda}$, while the
position of the NS5-branes in the confining vacuum scales as
$\sqrt{\lambda}$. Thus, it is clear just from looking at the 5-branes
that these vacua cannot have supergravity duals, since the position of
the 5-branes (which are sources for the supergravity fields) is not
fixed in the large $N$ limit, (nor in the large $N$, large $\lambda$
limit). We will see that the same result follows from looking at the
large $N$ limit of the chiral condensates.

Using our results above (using either the field theory or the results of  
\polstr, since both agree whenever supergravity might
be valid) we can easily compute the behavior of the chiral condensates
in the large $N$ limit. Let us start with the large $N$, fixed (large)
$\lambda$ limit. We find
in the Higgs vacuum $(p=N,q=1)$
\eqn\higgsvev{
\vev{\co_2}_{SG}\propto N^4 m^2/\lambda,}
%
%
while in the $k$'th confining vacuum
\eqn\confvev{\vev{\co_2}_{SG}\propto N^2m^2\lambda}
%
%
(with an additional term scaling as $k N m^2
\lambda^2$ if we keep $k$ constant and non-zero in the large $N$ limit).
Note
that $m$ is normalized so that it couples directly to a descendant of ${\cal
O}_2$, so we expect it to be exactly the natural parameter from the
supergravity point of view. 
%
%
Both expressions are too large and have the wrong $N$ or $\lambda$ scaling to
appear naturally in the constant $\lambda$ supergravity limit.
%
%
%
Similarly, the vacuum expectation value of $Q^2 \co_2$
in the zeroth confining vacuum includes terms scaling as $N^2 m^3
\lambda$ , while in the Higgs vacuum it scales as $N^4m^3/\lambda$,
both of which have the wrong $N$
or $\lambda$ scaling to appear in the constant $\lambda$ limit in
supergravity. A related observation made in \dorkum\ is that the
tension of a BPS domain wall which interpolates between Higgs and
confining vacua scales like $N^{4}$, which is hard to explain in terms
of a configuration interpolating between fixed SUGRA backgrounds. 
In contrast, this behavior is correctly 
reproduced by the construction of domain walls as five-brane 
junctions given in \polstr. Another puzzle raised in \dorkum\ is
that, in the context of the singular solutions of \gppz, 
there is no obvious explanation for the worldsheet instantons 
contributing to quantities in the confining vacuum. As discussed in
the previous section, this also has a natural resolution in the 
construction of \polstr.

In principle, even if the large $\lambda$ limit is not described by
SUGRA, it could still describe the limit of large $N$ and constant
$\tau$, which could correspond to a supergravity background at
some fixed value of the string coupling. (A weak coupling
expansion will generally not be possible in such a vacuum; 
for example, these could be vacua involving NS 5-branes, for which string
perturbation theory breaks down.) As described above, the solutions of
\polstr\ suggest that this is not possible in the Higgs and confining
vacua, and we can verify also that
the chiral VEVs in these vacua are too large for SUGRA in this limit;
both $\vev{\co_2}_{SG}$ and $\vev{Q^2 \co_2}_{SG}$
scale as $N^3$. In the Higgs vacuum we find $\vev{\co_2}_{SG}\simeq -N^3
m^2 \Im(\tau) / 96\pi$ and $\vev{Q^2 \co_2}_{SG}\simeq N^3 m^3 \Im(\tau)
/ 24\pi$, while in the zeroth confining vacuum (with $\theta_{YM}=0$) we
find $\vev{\co_2}_{SG} \simeq N^3 m^2/ (96 \pi \Im(\tau))$ and $\vev{Q^2
\co_2}_{SG} \simeq N^3 m^3/ (24\pi \Im(\tau)) $. These 
expressions are consistent with the $S$-duality
relation between these vacua (recall that $S$-duality also changes the
phase of $m$).

Thus, we conclude that the Higgs and confining vacua are not dual to 
fixed supergravity backgrounds in the large $N$ limit (even if we lift
the requirement of having a good string perturbation expansion around
these backgrounds). The only dual description of these vacua is through the
full solutions of \polstr. 

\newsec{Supergravity-like large $N$ limits}

As we saw above, the Higgs and confining vacua are not dual to fixed
supergravity backgrounds in the large $N$ limit; however, there is
another large $N$ limit of massive vacua which could be dual 
to such a background (at least asymptotically).
This is the limit of
large $N$, when we take the coupling to be constant and 
look at sequences of vacua with $p/q$ values converging to a fixed number in
the large $N$ limit. 
For example, we can take $p/q=2$ and look
only at values of $N$ of the form $N=pq=2q^2$ for integer $q$'s. If we
also take $k/q$ to be constant in the large $N$ limit (for example, we
can take $k=0$), we find in this limit that $\vev{\tilde{\co}_2}$ and
$\vev{Q^2 {\tilde\co}_2}$
converge to expressions scaling as $N^2$ , consistent with a supergravity
description. 
The expressions for the higher $\vev{\co_k}$'s which can be found in the 
appendix of \dorkum\ (without mixings) 
also appear to have the correct scaling in
this limit.

In the stringy description of the massive vacua in \polstr, the main
difference between these vacua and the confining and Higgs vacua is in
the position of the 5-branes. 
In the large $N$ limit with
constant $p/q$, the 5-branes are at a finite radial position in $AdS$ 
units. For example, for $\Im(\tt)\gg 1$ where we can describe these
vacua in terms of D5-branes, the radial position of the D5-branes in
units of the $AdS$ radius is a constant times $m\sqrt{\Im(\tt)}$. Thus,
these solutions converge to a fixed background of string theory in the
large $N$ limit. In terms of the solutions of \polstr\ the number of
5-branes grows in this limit but $\alpha'$ becomes smaller, since we
are keeping the string coupling fixed, such that the effect of the
5-branes on the background remains the same. Because of this property
it makes sense to compare this limit with the supergravity solutions for
deformations such as those of \gppz, and we will do this below. Note
that unlike $AdS_5\times S^5$, the vacua we describe here only exist
for specific values of $N$; e.g. if we take $p/q=2$ then $N$ has to be
of the form $N=2q^2$ for some integer $q$. However, this quantization
is invisible in the supergravity limit (though it is obvious when
these vacua are described in terms of branes \polstr).

In \gppz, solutions of five dimensional supergravity were described
that correspond to mass deformations of the type we analyzed here
(with equal masses). These solutions are all singular. However,  
since five dimensional supergravity is believed to be a
consistent truncation of type IIB supergravity on $AdS_5\times S^5$, it is
believed that 
they can be extended into solutions of ten dimensional supergravity,
which would presumably still be singular. The authors of \gppz\ 
looked only for solutions with
${\vev{\co_2}}_{SG}=0$, and found such solutions both with $\vev{Q^2
\co_2}_{SG}=0$ and with ${\vev{Q^2 \co_2}}_{SG}\neq 0$.
The generic vacuum that we find in the limit discussed above has both
${\vev{\tilde{\co}_2}}$ and ${\vev{Q^2 \tilde{\co}_2}}$ non-zero, so it does not
correspond to any of the vacua of \gppz.
However, for particular values of $\tau$ we can get
${\vev{\tilde{\co}_2}=0}$, and these values could correspond to some of the solutions of
\gppz\ (which necessarily exist for all $\tau$, since there
is no potential for the dilaton in the supergravity approximation). 
Using the conjectured form \expl, \nmixedvevs\ for the VEVs, we find that 
$\vev{\tilde{\co}_2}=0$ is solved by $\tt=e^{2\pi i/3}$ or values
related to this by $SL(2,\IZ)$. Plugging this value back into
\nmixedvevs, we find that these solutions also satisfy $\vev{Q^2
\tilde{\co}_2}=0$. Thus, we seem to have found candidate vacua that match the solutions 
of \gppz\ with $\vev{Q^2 \co_2}_{SG}=0$. Obviously this matching
would only be valid in a regime where the solutions of \gppz\ are not
singular, otherwise these solutions must be corrected. 
In fact, in the cases where the analysis of \polstr\ is
valid, the solutions can acquire corrections (due to the presence of
the 5-branes) even before the
singularity is reached, but always a finite distance away in $AdS$ units.
We conclude that the singularity found by \gppz\ in the solution with
$\vev{Q^2 \co_2}_{SG}=0$ should be resolvable in
string theory, and that similar resolvable singular solutions should
exist with non-zero $\vev{\tilde{\co}_2}$ and $\vev{Q^2 {\tilde\co}_2}$ related by
\nmixedvevs.

On the other hand, we find no solutions which in the large $N$ limit
have $\vev{\tilde{\co}_2}=0$ and $\vev{Q^2 {\tilde\co}_2}$ non-zero. Thus, it seems
that the singularities found in solutions with this property in \gppz\
would not be allowed in string theory, at least in asymptotically $AdS$ 
spaces, since the field theory does
not appear to have any corresponding vacua.

Our analysis is based on the expressions which we only
know to be true in the large $N$ limit. 
For finite $N$ we do not have good arguments in favor of supergravity
expressions coinciding precisely with the definition of the field
theory operators which is ${\tilde S}$-duality covariant. We might 
actually get a larger family of solutions with four real
parameters ($\tau$, $p/q$ and $k/q$), which degenerates into a
two-parameter family of solutions in the large $N$ limit. 
If this is the case then
in the full type IIB string theory there will be solutions 
with $\vev{\co_2}_{IIB}=0$ and
$\vev{Q^2 \co_2}_{IIB}$ non-zero, but $\vev{Q^2 \co_2}_{IIB}$ in these
solutions would 
have to grow slower
than the standard supergravity scaling in the large $N$ limit, so that
these solutions cannot correspond to those found in \gppz.


So far we have focused only on the massive vacua of the deformed
$\cn=4$ theory, even though for large $N$ there are many more massless
vacua (with massless photons and no mass gap). Unfortunately, the solutions
to the equations of motion of \tdsup\ 
corresponding to generic massless vacua are not known, and therefore their
analysis is much harder. 
It seems reasonable to expect that the operator VEVs
in the massless vacua will be of the same order as those in the
massive vacua, so that there may also be series of massless vacua that
converge in the large $N$ limit to SUGRA solutions. In principle they
could even converge to the same solutions, meaning that the
singularity in these solutions should have more than one possible
resolution in string theory. Examples of this are provided by  
$SL(2,\IZ)$ invariant massless vacua which exist for certain values of $N$. 
For example, when $N$ is of the form $N=(l^2-1)k^2/l^2$ for
integers $k$ and $l$ (with $k$ divisible by $l$), the superpotential  
\tdsup\ is extremized by $X_a = (i+j\tau)/k$, where $i$ and
$j$ go over all pairs of integers from $0$ to $k-1$ except those where
both integers are divisible by $l$. It is easy to check that this
vacuum is $SL(2,\IZ)$ invariant. Another series of
$SL(2,\IZ)$-invariant massless vacua
arises for $N=k(k+1)/2$ from the $N$ dimensional representation of
$SU(2)$ corresponding to blocks of size $(1,2,3,\cdots,k-1,k)$. Since
this is the only representation which leads to a $U(1)^{k-1}$ gauge
theory at low energies this vacuum must also be $SL(2,\IZ)$ invariant.
In such $SL(2,\IZ)$ invariant vacua we must have $W=0$ (since
$W/\tm_1 \tm_2 \tm_3$ must be a holomorphic modular form of weight
two, which does not exist), so they have $\vev{\co_2}=\vev{Q^2
\co_2}=0$ for the unmixed operators in the field theory, just like the massive
vacua with $p=q$ and $k=0$. The
VEVs of the mixed operators in these vacua are thus given by
\expl\ and \nmixedvevs, with $p=q$ and $k=0$. These vacua seem to
also converge to the SUGRA solution found in
\gppz\ with $\vev{Q^2 \co_2}_{SG}=0$.
The values of $N$ which give rise to these massless vacua 
are of course different from those 
that give rise to the massive vacua with $p=q$, but
in the large $N$ limit the solutions corresponding to these different
vacua would look very similar (although the string theory resolution of
the singularities is quite different, as in \polstr).

What do these results teach us about the resolution of the singularities
appearing in the solutions of \gppz\ ? Assuming that a complete analysis
of the massless vacua does not change our conclusions,
it appears that the singularity
appearing in the $\vev{Q^2 \co_2}_{SG}=0$ solutions should be resolvable
while the others should not. In other words, there should be no solution of
string theory in asymptotically $AdS$ space
which converges (for large enough radial coordinates) to
the singular solution with $\vev{Q^2\co_2}_{SG}\neq 0$. 
There seems to be no obvious way of distinguishing the 
resolvable singularities  
from the others directly.
Of course, we expect that whenever a singularity can be
resolved it will be in terms of some brane configuration as in
\polstr, so the claim here is that no brane configuration can resolve
most of the singularities of \gppz\ in the sense described above.
Note that all the singularities of \gppz\ obey the criteria of Gubser
\gubser\ for ``good'' singularities, so we seem to have an example of
``good'' singularities which are still disallowed. In this case it
seems that the finite temperature criterion of \gubser\ should not be
relevant, since we would not expect the solution of a particular
vacuum to have a finite temperature generalization; rather, at a
temperature of order $m$ (where we might expect to see a smooth
horizon), the field theory is presumably in some state
which is a superposition of the different vacua (and of other states).

It is not clear if this analysis teaches us something about
resolving singularities with continuous parameters in flat space, such
as those that appeared in \refs{\kssone,\adks,\ksstwo}. There is
obviously no direct relation since we are discussing ten dimensional
singularities in asymptotically $AdS$ spaces, 
while they discuss five dimensional singularities in flat space.
However, our results suggest that
generically it would not be possible to resolve such singularities in
string theory.

\vskip 1cm
{\bf Acknowledgements:}
O. A. would like to thank K. Intriligator, J. Polchinski and 
M. Strassler for
useful discussions, and to thank SLAC, the University
of Chicago and the Caltech/USC center for theoretical physics for
hospitality during the course of this work.
S. P. K. would like to thank N. Kaloper, K. Konishi and V. Periwal for useful
discussions. The work of O. A. was supported in part by DOE grant
DE-FG02-96ER40559. 
S. P. K. acknowledges support of DOE grant DE-FG03-96ER40956. 
N. D. acknowledges the support of TMR network grant FMRX-CT96-0012 
and of a PPARC Advanced Research Fellowship. 

\listrefs

\end